\documentstyle[preprint,floats,aps,psfig]{revtex}
\tightenlines
\begin{document}

\draft

\title{ Effect of form factors in fits to photoproduction data}
\author{R. M. Davidson\thanks{davidr@rpi.edu} }
\address{Department of Physics, Applied Science and Astronomy \\
         Rensselaer Polytechnic Institute, Troy, New York 12180-3590}
\author{Ron Workman\thanks{rworkman@gwu.edu} }
\address{Center for Nuclear Studies and Department of Physics \\
        The George Washington University Washington, DC 20052}

\draft
\date{\today}
\maketitle

\begin{abstract}

We compare the effects of several form factor recipes in 
fits to pseudoscalar meson photoproduction data. The specific
examples of pion and kaon photoproduction are used to illustrate 
how different choices can alter the results of such analyses.

\end{abstract}
\vspace*{0.5in}


The study of meson photo- and electroproduction is
enjoying a resurgence, due mainly to a program of
precise measurements at new facilities in the U.S. and
Europe. One important goal of these studies is 
an improved understanding of the baryon resonances and their
photo-decay amplitudes. It is also hoped that these new
experiments and subsequent analyses will reveal states not found in
previous fits to pion induced and pion production reactions. 
 
In order to obtain resonance information from cross sections and
polarization measurements, one either performs a multipole
analysis, followed by a separation of resonance/background
contributions, or directly fits data in terms of a model
explicitly containing resonance parameters. In the absence of
complete experimental information, some theoretical input is
always necessary. The use of a truncated partial-wave
series, for example, simplifies the problem considerably, but requires
a model for the high partial waves. In the resonance region,
these high partial waves have generally been taken from
a Born approximation.

In some cases, particularly in pion photoproduction, a Born approximation
can be justified, at least qualitatively, through comparisons
with low-energy multipole amplitudes and cross sections. A number
of low-energy s- and p-wave multipoles are remarkably similar to
those generated using a (unitarized) Born approximation. The
use of a Born approximation for higher partial waves also seems
to account for the forward peaking seen in charged-pion 
production.

While the use of a Born approximation is helpful at low energies,
serious problems arise as the energy increases. In particular,
the use of a pseudovector coupling scheme in pion photoproduction
is advantageous near
threshold but diverges badly compared to data
above the delta resonance region.
In the Mainz analysis\cite{mainz}, 
this problem was handled by allowing the
$\pi NN$ coupling to be a mix\cite{mix} of pseudoscalar and pseudovector
terms, becoming purely pseudoscalar at higher energies.
In the kaon photoproduction case, however, even the pseudoscalar
scheme diverges over the region of interest, and several attempts
have been made to introduce a cutoff scheme for
Born contributions\cite{tkb,ohta,workman,haberzettl,DW}.

In order to compare the different approaches, we write the charged
pseudoscalar meson photoproduction amplitude off protons in terms of the
usual four gauge invariant amplitudes,
\begin{equation}
\epsilon \cdot M_{fi} = \bar u_f \sum_{j=1}^4 A_j M_j u_p \; ,
\end{equation}
with the explicitly gauge invariant representation
\begin{eqnarray}\label{invamp}
M_1 & = & -\gamma_5 \epsilon \cdot \gamma k\cdot \gamma \; ,\\
M_2 & = & 2\gamma_5 ( \epsilon \cdot p_1 k\cdot p_2
                       - \epsilon \cdot p_2 k\cdot p_1 ) \nonumber \; , \\
M_3 & = & \gamma_5 ( \epsilon \cdot \gamma k\cdot p_1
                       - \epsilon \cdot p_1 k\cdot \gamma ) \nonumber \; , \\
M_4 & = & \gamma_5 ( \epsilon \cdot \gamma k\cdot p_2
                       - \epsilon \cdot p_2 k\cdot \gamma ) \nonumber \; ,
\end{eqnarray}
where $k$ and $q$ give the photon and pion four-momenta, and $p_1$ and $p_2$
are the respective initial and final baryon four-momenta.
The pseudoscalar Born terms are given by
\begin{eqnarray}\label{as}
A_1 & = & { {GeF(s,m_f^2 , \mu^2)}\over {s-m^2}} (1+\kappa_p) +
{{GeF(m^2 ,u, \mu^2)}\over {u-m_f^2}} \kappa_f \; , \\
A_2 & = & {{2Ge}\over {(s-m^2)(t-\mu^2)} } \hat F \; , \nonumber \\
A_3 & = & {{Ge}F(s, m_f^2 , \mu^2 )\over {s-m^2}} {\kappa_p \over m} \; ,
\nonumber \\
A_4 & = & {{Ge}F(m^2 ,u, \mu^2)\over {u-m_f^2}} {\kappa_f \over m} \; ,
\nonumber
\end{eqnarray}
where $m$ is the mass of the proton, $\mu$ is the mass of the produced
meson, $\kappa_p$ is the anomalous magnetic moment of the proton,
$m_f$ is the mass of the final baryon, and $\kappa_f$ is the
anomalous magnetic moment of the final baryon. For the strong coupling
constant $G$, in the case of pion production, we
take $G$ = $\sqrt{2}G_{\pi^0 pp}$ with $G_{\pi^0 pp}^2 /(4\pi)$ = 13.75.

In eq. (\ref{as}),  $F(s,u,t)$ is a form factor modifying the $s-$,
$u-$, or $t-$channel
exchanges and the factor $\hat F$ is constructed to modify $A_2$ in a way such
that any gauge-invariance-violating term can be cancelled by an
additional contact term.
The point-like Born terms
are obtained by setting all form factors equal to unity.
In the minimal substitution scheme of Ohta\cite{ohta},
$\hat F = 1$ and the $A_2$ amplitude is unmodified.
In Ref.\cite{haberzettl}, the functional form of
$\hat F$ was taken to be
\begin{equation}\label{hfhat}
\hat F = a_1 F_1(s) + a_2 F_2(u) + a_3 F_3(t) \; ,
\end{equation}
where
\begin{eqnarray}
F_1 (s) &=& F(s, m_f^2 , \mu^2) \; , \nonumber \\
F_2 (u) &=& F(m^2 , u , \mu^2) \; , \nonumber \\
F_3 (t) &=& F(m^2 , m_f^2 , t) \; ,
\end{eqnarray}
and $\hat F$ was
subject only to the constraint $a_1 + a_2 + a_3 = 1$. In Ref.\cite{DW}, it was
shown that Haberzettl's form is unacceptable except at the soft photon point,
and a much
more restrictive form was found which is in all cases consistent
with the pole structure of the amplitude.
For charged-meson photoproduction, one allowed
form for $\hat F$ is
\begin{equation}\label{fhat}
\hat F = F_1 (s) + F_3 (t) - F_1 (s)F_3 (t) \; ,
\end{equation}
which we will use to compare with previous works.

The Born terms can, in principle, give a
significant contribution to charged-meson photoproduction, 
and their modification could alter the findings for associated 
resonance contributions. For example,
in kaon photoproduction 
the divergence of a point-like-Born contribution to the 
total cross sections is well known, and attempts to
cure this problem, by modifying the
individual multipoles, were described in Ref.\cite{tkb}.
Ohta's prescription, while damping all but amplitude $A_2$, still
gives a Born contribution which grows too rapidly. 
In Ref.\cite{DW} it was shown that the $\hat F$ form given in
Eq.~6 successfully damps the Born contribution to the total cross
section, and thus could be used in a phenomenological analysis of the
photokaon data. It is difficult to judge how a proper treatment of
$\hat F$ might affect the results of such an analysis, but we note
there is a broad kinematic region where $\hat F$
is numerically close to $F_3 (t)$ when we use standard 
expressions for the form factors;
\begin{eqnarray}\label{ff}
F_1 (s) &=& {\Lambda^4 \over \Lambda^4 +(s-m^2 )^2} \; , \nonumber \\
F_2 (u) &=& {\Lambda^4 \over \Lambda^4 +(u-m_f^2 )^2} \; , \nonumber \\
F_3 (t) &=& {\Lambda^4 \over \Lambda^4 +(t-\mu^2 )^2} \; .
\end{eqnarray}
Thus, our $\hat F$ is numerically close to that used
in Ref.\cite{haberzettl}, i.e., 
$\hat{F} \approx F_3(t)$ corresponding to a particular
choice of their weight coefficients. However, this choice was not
continued in subsequent works\cite{feuster,D13etc}.

If form factors are important in the case of kaon production, one must
consider their potential effect in pion production as well. This is of
particular importance since most of our current knowledge of electromagnetic
decays of nucleon resonances has been extracted from analyses of pion
photoproduction data. Most of these analyses have used point-like
Born terms which might bias the extracted resonance parameters.
As in kaon production,
the total cross section generated by the pseudoscalar Born terms for
$\pi^+ n$ production exceeds the experimental value starting at
a photon lab-energy ($E_{\gamma}$) of about 450 MeV. Putting in form factors,
as above, is one way to cure this problem. However, the effect of these
form factors on the differential cross section is cause for concern.
The experimental cross section exhibits a forward peak which is usually
associated with the Born terms. With the addition of form factors, 
the Born terms may lose this forward peaking behavior.
The peak seen in forward $\pi^+ n$ cross sections is a result
of destructive interference amongst the small-$l$ multipoles and the
large-$l$ multipoles generated by the $A_2$ amplitude
in Eq. (\ref{as}), which in a specific gauge
can be interpreted as the $t$-channel pion exchange contribution.
In many cases, the small-$l$ multipoles are
fitted to the data while the large-$l$ waves are assumed to be given by the
point-like
Born terms. In other cases, the full point-like
Born term is added to phenomenological resonance contributions.
The effect of form factors in the fitting process is expected to be
sensitive to this choice. 
Let us first consider how 
form factors modify the large-$l$ multipoles.

Note
that large-$l$ multipoles are dominated by the $A_2$ amplitude and,
ignoring irrelevant factors,
to obtain the multipoles, one needs integrals of the form
\begin{eqnarray}
M_l = {1 \over 2} \int_{-1}^1 {P_l (x) \over t-\mu^2} \left[
F_1 (s) +F_3 (t) -F_1 (s) F_3 (t) \right] {\rm d}x \; ,
\end{eqnarray}
where $x$ is the cosine of the center-of-momentum (cm) scattering angle
and Eq.~\ref{fhat} has been used for $\hat F$.
Now define
\begin{equation}
M_l^0 = {1 \over 2} \int_{-1}^1 { P_l (x) \over t-\mu^2}{\rm d}x
= - {1 \over 2qk} Q_l (z) \; ,
\end{equation}
with $z$ = $\omega /q$ where $\omega$ (q) is the pion energy (three-momentum)
in the cm frame.
In the absence of form factors, $M_l^0$ would give the
point-like result. Using our forms of $\hat{F}$ and $F_3 (t)$, we obtain
\begin{eqnarray}
M_l =  M_l^0 -I_l (1-F_1 (s)) \; ,
\end{eqnarray}
where
\begin{eqnarray}
I_l &=& { 1 \over 2} \int_{-1}^1 { P_l (x) ( t-\mu^2) \over \Lambda^4 +
( t-\mu^2)^2 } {\rm d}x \; , \nonumber \\
 &=& - { 1 \over 2qk } {\rm Re}Q_l (z_{\Lambda}) \; ,
\end{eqnarray}
with
\begin{equation}
z_{\Lambda} = {\omega \over q } - { i\Lambda^2 \over 2qk } \; ,
\end{equation}
and $k$ is the energy of the photon in the cm frame.

To check the relative importance of $M_l^0$ and $I_l$, we make use of
the series expansion in $1 / x$ for $Q_l (x)$, the leading term
being proportional to $1 / x^{l+1}$.
The usefulness of this series in
evaluating $Q_l (x)$ depends on the particular value of $x$, but the important
point to notice is that every term in the expansion for $Q_l (z_{\Lambda})$
is suppressed by at least
\begin{equation}
\left( { z \over | z_{\Lambda}| }  \right) ^{l+1}
\end{equation}
compared to the same term in the expansion of $Q_l (z)$, where
\begin{equation}
|z_{\Lambda}| = \sqrt{ {\omega^2 \over q^2}+{\Lambda^4 \over 4q^2
k^2 }} > z \; .
\end{equation}
Thus, for sufficiently large $l$, $I_l$ is negligible compared to
$M^0_l$.  The $u$-channel can also
contribute to the large-$l$ multipoles.
However, this contribution is controlled
by $Q_l (z_n )$ where $z_n$ = $E_f /q$, $E_f$ being the final
baryon energy in the cm frame. As $z_n$ $>$ $z$, one sees that the $u$-channel
contribution is negligible compare to the $t$-channel contribution for
sufficiently large $l$.

The usefulness of this argument depends on how large 
$l$ must be before $M_l^0$ dominates.
This depends of the value of $\Lambda^2 /2qk$
because when $s$ $\rightarrow$ $\infty$, $z_{\Lambda}$ $\rightarrow$
$z$. However, as most multipole analyses are restricted to $E_{\gamma}$
$<$ 2 GeV, we concentrate on energies in this range. For $l$ = 0
and 1, form factors significantly affect the multipoles at almost all
energies.
For $l$ = 2, form factor effects set in at about $E_{\gamma}$ =
500 MeV, and with increasing $l$, form factors effects set in at higher
$E_{\gamma}$.

Although the large $l$ multipoles
are given very accurately by the Born terms without form factors, it could be
the case that these large $l$ multipoles are unimportant. To check this,
we show in Fig.~1 a comparison of the current SAID solution (solid line)
with the experimental data \cite{pindata}
for differential cross sections at $E_{\gamma}$ =
1002 MeV. To determine the importance of the $l$ $\ge$ 6 multipoles, we
calculated the differential cross section using multipoles up to $l$ = 5
(dashed line). Evidently, the $l$ $\ge$ 6 multipoles are important in
reproducing the forward peak. As a verification that the Born terms
with form factors are capable of reproducing this peak, we have taken
our amplitude for the Born terms with form factors, subtracted off their
$l$ $\le$ 5 multipoles, and added the rest to the SAID $l$ $\le$ 5 multipoles.
The resulting differential cross section is shown as the dotted line, which
is nearly identical to the solid one. The full Born term (dashed-dotted)
is also plotted
with form factor modification, to show how the changes in lower partial
waves affect the forward peaking behavior.

The results in Fig.~1 demonstrate that with a reasonable cutoff
it is possible to damp the Born term contribution while at the
same time not damp the high-$l$ multipoles which are needed to
produce the forward peak in the 
cross section. While results will vary depending
on what one chooses for the form factors, it is true in general that
the high-$l$ multipoles will be dominated by the nearest singularity
in $x$. This means that the high-$l$ multipoles will be given by the
point-like-Born contribution unless one uses form factors which have
a singularity in $x$ nearer than the one arising from the pion pole.
Given these results, our conclusion is that multipole analyses of the type
given in Ref.\cite{said}
are on firm ground in the sense that if Born terms with
form factors were to be used in the analysis, the a priori unknown
parameters in the $l$ $\le$ 5 multipoles, which are fitted to data, would
readjust to give a multipole solution very close to the present one.
On the other hand, it is less clear how these form factors would influence
results based on effective Lagrangian approaches\cite{dmw} or
isobar models\cite{mainz}. If the
analysis is restricted to the first resonance region, we expect no serious
problem and, indeed, results for the Delta(1232) parameters obtained from
various approaches are in good agreement \cite{rpivpi}. However, if the
analysis is performed over a larger energy region, problems may arise.
At higher energies, the point-like-Born contribution must be partially
cancelled by some mechanism, i.e., the resonance or vector meson exchange
contributions, which could bias the extracted resonance parameters. We note
that simply using Born terms with form factors is not the solution to
this problem and might actually worsen the problem. If this were done, the
small-$l$ multipoles, which are often resonance dominated, would have
to readjust in order to produce the forward peak in the cross section, and
thus, there is again the possibility of bias in the extracted resonance
parameters.

\acknowledgments

This work was supported in part by the U.~S. Department of 
Energy Grants DE--FG02--99ER41110 and DE--FG02--88ER40448.
R.W. gratefully
acknowledges a contract from Jefferson Lab under which
this work was done.  Jefferson Lab is operated by the 
Southeastern Universities Research Association under the
U.~S.~Department of Energy Contract DE--AC05--84ER40150.

\eject


\begin{figure}
\caption{Differential cross section for $\gamma p$ $\rightarrow$ $\pi^+ n$
versus angle at $E_{\gamma}=1002$ MeV. Plotted are the SAID
\protect\cite{said} fit (solid), SAID fit minus waves beyond
$l = 5$ (dashed), and SAID fit with a replacement of $l\ge 6$
waves with form-factor
modified Born contribution (dotted). Also shown is the Born
contribution with form factors (dashed-dotted).
Data from Ref.\protect\cite{pindata}.}
\end{figure}

\vfill
\eject


\begin{thebibliography}{99}

\bibitem{mainz}
D. Drechsel, O. Hanstein, S.S. Kamalov, and L. Tiator,
Nucl. Phys. {\bf A645}, 145 (1999).

\bibitem{mix}
This approach is supported by the study of
S. Kondratyuk and O. Scholten,
Phys. Rev. C {\bf 59}, 1070 (1999).


\bibitem{tkb}
H. Tanabe, M. Kohno, and C. Bennhold,
Phys. Rev. C {\bf 39}, 741 (1989).

\bibitem{ohta}
K. Ohta,
Phys. Rev. C {\bf 40}, 1335 (1989).

\bibitem{workman}
R.L. Workman, H.W.L. Naus, and S.J. Pollock,
Phys. Rev. C {\bf 45}, 2511 (1992).

\bibitem{haberzettl}
H. Haberzettl, C. Bennhold, T. Mart, and T. Feuster,
Phys. Rev. C {\bf 58}, R40 (1998).

\bibitem{DW}
R.M. Davidson and R.L. Workman,
Phys. Rev. C, in press.

\bibitem{feuster}
T. Feuster and U. Mosel, Phys. Rev. C {\bf 59}, 460 (1999).

\bibitem{D13etc}
T. Mart and C. Bennhold,
Phys. Rev. C {\bf 61}, 012201(R) (1999);
T. Mart,
Phys. Rev. C {\bf 62}, 038201 (2000);
F.X. Lee, T. Mart, C. Bennhold, H. Haberzettl, L.E. Wright,
nucl-th/9907119.

\bibitem{pindata}
Data from the SAID database \hbox{(http://gwdac.phys.gwu.edu)}. This
source also contains representative fits from several different
groups.


\bibitem{said}
R.A. Arndt, I.I. Strakovsky, and R.L. Workman,
Phys. Rev. C {\bf 53}, 430 (1996).

\bibitem{dmw}R.M. Davidson, N.C. Mukhopadhyay and R.S. Wittman, Phys. Rev.
D {\bf 43}, 71 (1991).


\bibitem{rpivpi} R.~M.~Davidson, Nimai C.~Mukhopadhyay, M.~S.~Pierce,
R.~A.~Arndt I.~I.~Strakovsky and R.~L.~Workman,
Phys. Rev. C {\bf 59}, 1059 (1999).


\end{thebibliography}
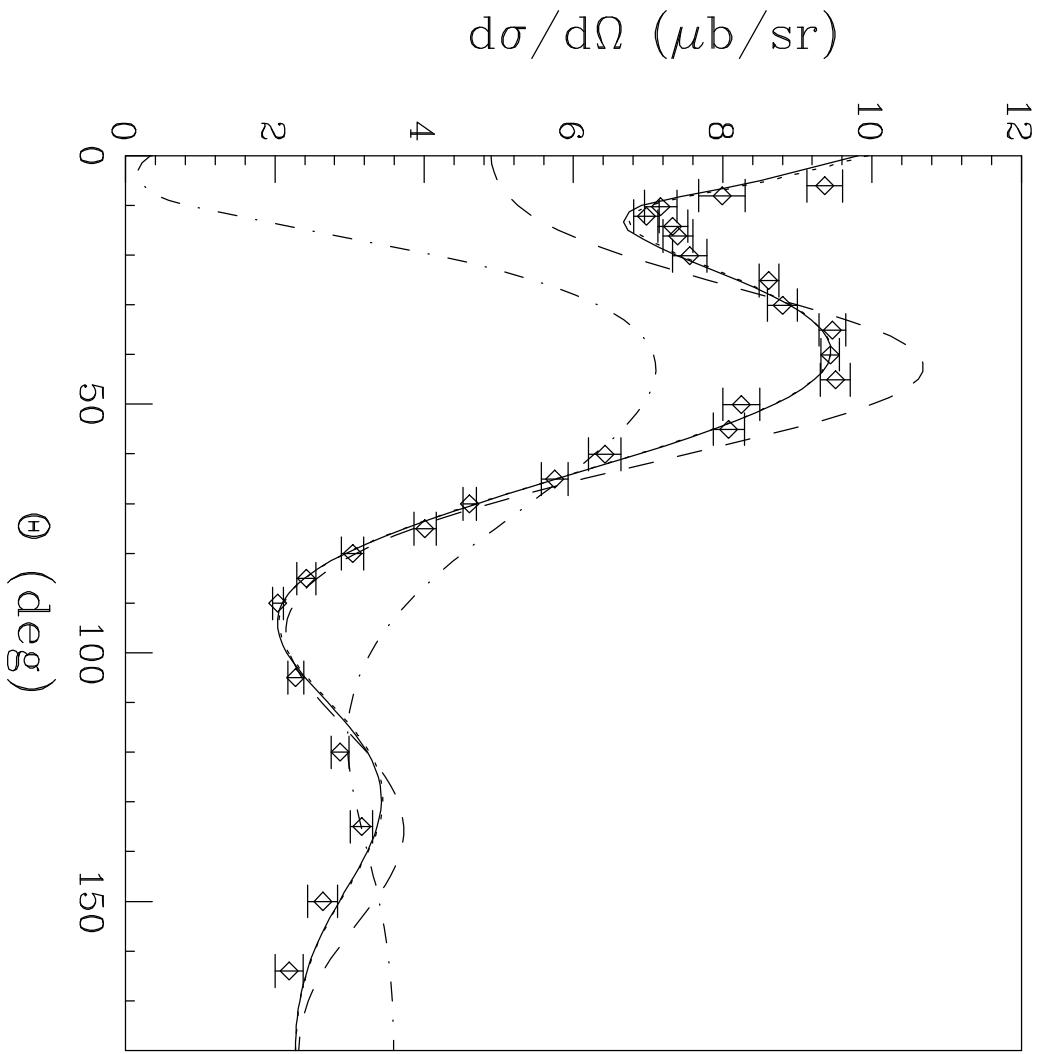
\end{document}